\documentclass[prc,showpacs,twocolumn]{revtex4}
\usepackage{amssymb}
\usepackage{graphicx}
\usepackage{dcolumn}
\usepackage{bm}
\usepackage{multirow}

\begin{document}

\title{Pairing effect in thermal shape fluctuation model on the width of giant dipole resonance }
\author{A.K. Rhine Kumar$^1$}
\email{rhinekumar@gmail.com}
\author{P. Arumugam$^1$}
\author{N. Dinh Dang$^2$}
\affiliation{$^1$Department of Physics, Indian Institute of Technology Roorkee,
Uttarakhand - 247 667, India\\
$^2$RIKEN Nishina Center for Accelerator-Based Science, RIKEN 2-1 Hirosawa, Wako city, 351-0198 Saitama, Japan,
and Institute for Nuclear Science and Technique, Hanoi, Vietnam.}
\date{\today}

\begin{abstract}
We present an approach that includes temperature-dependent shell effects and fluctuations of the pairing field in the thermal shape fluctuation model (TSFM). We apply this approach to study the width of giant dipole resonance (GDR) in $^{97}$Tc, $^{120}$Sn and $^{208}$Pb.  Our results demonstrate that the TSFM that includes pairing fluctuations can explain the  recently observed quenching in the GDR width. We also show that to validate pairing prescriptions and the parameters involved, we require more and precise data.
\end{abstract}

\pacs {24.30.Cz,  21.60.-n,  24.60.-k}

\maketitle

\section{Introduction}
The study of nuclear properties at high temperature, spin and
isospin has gained much of interest in recent times. Thanks to the recent
developments in experimental facilities, these highly-excited nuclei are
becoming more accessible and provide   theorists with a challenging task.
Apart from these extremes, there are still some unexplored regimes of hot
nuclei. The properties of nuclei at very low temperatures and the phase
transitions associated with that belong to such area where conclusive experimental results are scarce. At such low temperatures, the shell (quantal) and pairing effects are quite active though being modified by thermal effects. Among the famous and open questions in this regime are the existence of  pairing phase transition, the order of it if it exists, the role of fluctuations, etc. In hot nuclei, thermal fluctuations are expected to be large since the nucleus is a tiny finite system. Thermal shape fluctuations  and fluctuations in the pairing field are the dominating fluctuations and they have been so far studied separately within different models \cite{ALHAT,aruprc1,MORET40B,Dang044303,Dang064315,Dang054324re}. Both of these fluctuations are expected to be present at low temperatures. However, the interplay between them has not been investigated so far. The present work addresses this subject and we study the influence of this interplay on the experimental observables, namely the width of giant dipole resonance (GDR).

GDR is a fundamental mode of excitation of nuclei caused by the out of phase oscillations between the proton and neutron fluids under the influence of the electromagnetic field induced by the emitted/absorbed photon. In general, the resonance parameters of any resonating object  are related to the geometry of the  object.  In this way, the GDR width and its cross-section could yield direct information about the shape of the nucleus. This is only a macroscopic description of GDR and there are microscopic approaches, which couple the GDR to particle-hole, particle-particle and hole-hole excitations \cite{DangPLB,Dang:636,Dang:645}. For hot nuclei, which are not accessible by the discrete $\gamma$-ray spectroscopy and other conventional techniques, the measurement of GDR is considered to be a major probe to obtain the details of nuclear structure. The importance of measuring the GDR width  at low temperatures was insisted by one among the authors \cite{Dang:531} and the first results  were reported in Ref.~\cite{Heck43}, where it was found that the GDR width in $^{120}$Sn at $T=$ 1 MeV is nearly the same as that in the ground state. This data point was successfully explained only after treating properly the pairing correlations with
the phonon damping model (PDM) \cite{Dang044303,Dang:064319,Dang:014304}. Similarly, it was found in Ref.~\cite{camera155} that the GDR width in $^{179}$Au extracted at $T=$ 0.7 MeV is almost the same as its ground state value. However, it was misattributed to  the shell effects as in the case of $^{208}$Pb. This was clarified later \cite{aruepl} where the proper inclusion of shell effect was found to act in the opposite direction to raise the width and the pairing fluctuations were speculated to explain this anomaly. Preliminary results in this regard were reported in Refs.~\cite{aruriken,Dangarx,Rhine:AIP} where the importance of considering pairing in the thermal shape fluctuation model has been emphasized. The recent low temperature GDR measurements done at Variable Energy Cyclotron Centre, Kolkata \cite{mukhu9,Pandit2012434,pandit044325} highlight the interesting features of the GDR width in heavy nuclei observed at low temperatures.

The thermal shape fluctuation model (TSFM) \cite{aruprc1,ALHA,Kusn98,Orma217}, which is often used by experimentalists, describes the increase of the GDR width with temperature by averaging the GDR cross section over all the quadrupole shapes. However this model is known to largely overestimate the GDR width
in open-shell nuclei at low temperatures. The success of a proper treatment
of pairing within the PDM \cite{Dang044303,Dang:064319,Dang:014304,Balaram,Dang044333} suggests the necessity of including pairing correlations to cure this shortcoming of the TSFM. The PDM is a microscopic model, whose mechanism is different from that of the TSFM. The pairing has never been taken into account within the TSFM so far because of the (incorrect) assumption that the pairing gap disappears at $T\sim$ 1 MeV. Given the popularity of the TSFM the inclusion of pairing in the TSFM is quite important.  This is done for the first time in the present work. 

Apart from the TSFM, two phenomenological parameterizations have been reported \cite{Kusn98,Pandit2012434} which are very successful in approximating global behaviour of the GDR widths. In recent literature \cite{Pandit2012434,Balaram,pandit054327,pandit044325} these parameterizations referred to as, phenomenological TSFM (pTSFM) \cite{Kusn98}  and its modification to take into account the quenching of width at low $T$  has been referred to as a critical temperature included
fluctuation model (CTFM) \cite{Pandit2012434}. It has to be noted that these empirical formulae constructed to mimic the results of TSFM should not be confused with the TSFM itself. The pTSFM and CTFM are merely a phenomenological parameterizations based on empirical data, which has neither  microscopic nor macroscopic foundation.

It is interesting to note, although both the PDM and CTFM give results consistent with the measurements of the GDR width in $^{97}$Tc \cite{Balaram}, these two models are quite different from the TSFM. It is indeed mentioned in Ref.~\cite{Balaram} that it would be interesting to compare the data with TSFM by including the effect of thermal pairing.

Here we employ the thermal fluctuation model built on Nilsson-Strutinsky calculations with a macroscopic approach to GDR and examine the inclusion of the fluctuations in the pairing field. In Refs.~\cite{ansari011302,ansari024310} the same Hamiltonian was used to generate the free energy as well as the GDR observables in a consistent way, which led to a slow thermal damping of GDR width when compared to the experimental results.  The formalism adopted in the present work is well tested to reproduced several GDR observations at higher temperatures \cite{aruprc1,aruepl,aruepja}. This formalism is extended to include thermal pairing and our results are discussed in the forthcoming sections.

\section{Theoretical Framework}
The present theoretical approach is explained below in threefold within the models for A) deformation energy calculations, B) relating the shapes to GDR observables and C) considering the fluctuations due to thermal effects in finite systems.

\subsection{Deformation energy calculations \label{sec_defene}}
Here we follow the finite temperature Nilsson-Strutinsky method \cite{aruprc1}. The total free energy ($F_{\mathrm{TOT}}$) at a fixed deformation is calculated using the expression
\begin{equation}
F_{\mathrm{TOT}}=E_{\mathrm{LDM}}+\sum_{p,n}\delta F\;.  \label{FTOT}
\end{equation}%
The liquid-drop energy ($E_{\mathrm{LDM}}$) is calculated by summing up the
Coulomb and surface energies  corresponding to a triaxially deformed shape defined by the deformation parameters $\beta $ and $\gamma $. The shell correction ($\delta F$) is obtained with exact temperature dependence \cite{aruprc1}  using the single-particle energies given by the triaxial Nilsson model. While considering the pairing fluctuations, the nucleus is assumed to behave as a grand canonical ensemble (GCE), which allows fluctuations in particle number \cite{MORET}. The corresponding free energy is determined as 
\begin{equation}\label{free energy}
F=\left\langle H_{0}\right\rangle -\lambda N-TS\ ,
\end{equation}%
where $H_{0}$ is the nuclear hamiltonian which is independent of
temperature, $\lambda $ is the chemical potential, $N$ is the particle
number, $T$ is the temperature and $S$ is the entropy. The above
expression can be expanded to%
\begin{equation}
F=\sum_{i}(e_{i}-\lambda -E_{i})-2T\sum_{i}\ln [1+\exp (-E_{i}/T)]+\frac{%
\Delta ^{2}}{G}
\end{equation}
where $e_{i}$ are the single-particle energies obtained by diagonalizing $H_{0}$ with a harmonic oscillator basis comprising the first 12 major shells, $E_{i}=\sqrt{(e_{i}-\lambda )^{2}+\Delta ^{2}}$ are the quasiparticle energies, $\Delta $ is the pairing gap obtained by solving the temperature dependent BCS equations \cite{BCS57} by assuming a constant pairing
strength given by \cite{begston} $G_{p,n} =  [19.2 \pm 7.4(N-Z)]/A^2$. 
The smoothed free energy, in the Strutinsky way, can
be written as
\begin{eqnarray}
\widetilde{F}&=&2\sum_{i}(e_{i}-\lambda )\widetilde{n}_{i}-2T\sum_{i}\widetilde{s%
}_{i}\nonumber\\
&&+2\gamma _{s}\int_{-\infty }^{\infty }\widetilde{f}(x)x%
\sum_{i}n_{i}(x)dx - \frac{\Delta ^{2}}{G}
\end{eqnarray}
with the third term included to give better plateau conditions~\cite{aruprc1}.
Here $\widetilde{f}(x)$ is the averaging function given by%
\begin{equation}
\widetilde{f}(x)=\frac{1}{\sqrt{\pi }}\exp
(-x^{2})\sum_{m=0}^{p}C_{m}H_{m}(x)\;;  \label{FTIL}
\end{equation}%
$C_{m}=(-1)^{m/2}/(2^{m}(m/2)!)$ if $m$ is even and $C_{m}=0$ if $m$ is odd;
$x=(e-e_{i}^{\omega })/\gamma _{s}$, $\gamma _{s}$ is the smearing parameter
satisfying the plateau condition $d\widetilde{F}/d\gamma _{s}=0$; $\;p$ is
the order of smearing and $H_{m}(x)$ are the Hermite polynomials. The
averaged occupation numbers and single-particle entropies are given by $\widetilde{n}_{i}=\int_{-\infty }^{\infty }\widetilde{f}(x)\;n_{i}(x)\;dx$ and $\widetilde{s}_{i}=\int_{ -\infty }^{\infty }\widetilde{f}(x)\;s_{i}(x)\;dx$, respectively. The quasiparticle occupation numbers at finite temperature $T$ is given by
$n_{i}^{T}=\left[1+\exp (E_{i}/T)\right]^{-1}\;$, so that the total entropy can be written as%
\begin{eqnarray}
S&=&2\sum_{i}s_{i} \nonumber \\
&=&-2\sum_{i}\left[ n_{i}^{T}\ln n_{i}^{T}+(1-n_{i}^{T})\ln
(1-n_{i}^{T})\right].
\end{eqnarray}
For calculations without pairing ($\Delta=0$), we consider the canonical ensemble (CE) for which the expression for free energy reduces to those given in Ref.~\cite{aruprc1}.
 
\subsection{Nuclear shapes and GDR observables}
The nuclear shapes are related to the GDR observables using a model \cite{aruprc1,THIA1,THIA2} comprising an anisotropic harmonic oscillator potential with
separable dipole-dipole interaction. In this formalism the GDR Hamiltonian
can be written as
\begin{equation}
H=H_{osc}+\eta \ D^{\dagger }D\ +\chi \ P^{\dagger }P\ ,
\label{GDRhamiltonian}
\end{equation}
where $H_{osc}$ stands for the anisotropic harmonic oscillator hamiltonian,
the parameter $\eta$ characterizes the isovector component of the neutron and proton average fields and $\chi $ denotes the strength of the pairing interaction. The pairing interaction changes the oscillator frequencies [$\omega _{\nu }^{osc}$($\nu=x,y,z$)], resulting in the new set of frequencies $\omega _{\nu }=\omega _{\nu }^{osc}-\chi \omega ^{P}$, where $\omega ^{P}=\left(\frac{Z\Delta _{P}+N\Delta _{N}}{Z+N}\right)^{2}$ with $\chi$ having the units of MeV$^{-1}$. Alternatively, the  role of pairing is to renormalize the dipole-dipole interaction strength such that, $\eta=\eta_0-\chi_0\sqrt T\omega^P$, with $\chi_0$ having the units of MeV$^{-5/2}$.

Including the dipole-dipole and pairing interactions, the GDR frequencies in the laboratory frame are obtained as
\begin{equation}
\widetilde{\omega }_{z}=(1+\eta )^{1/2}\omega _{z}\;,
\end{equation}
\begin{eqnarray}
\widetilde{\omega }_{2,3} = && \left\{(1+\eta )\frac{\omega
_{y}^{2}+\omega _{x}^{2}}{2} \right. \nonumber \\ && \left.\pm\frac{1}{2}\left\{ (1+\eta )^{2}(\omega _{y}^{2}-\omega
_{x}^{2})^{2}\right\}^{1/2}\right\}^{1/2}.
\end{eqnarray}
With the pairing field, the GDR Hamiltonian has to be redefined, which will  affect the oscillator frequencies. The GDR cross section is constructed as a sum of Lorentzians given by
\begin{equation}
\sigma (E_{\gamma })=\sum_{i}\frac{\sigma _{mi}}{1+\left( E_{\gamma
}^{2}-E_{mi}^{2}\right) ^{2}/E_{\gamma }^{2}\Gamma _{i}^{2}}
\label{Eq.Cross}
\end{equation}
where $E_{m}$, $\sigma _{m}$ and $\Gamma $ are the
resonance energy, peak cross-section and full width at half maximum,
respectively. Here $i$ represents the number of components of the GDR and is determined from the shape of the nucleus \cite{THIA1,THIA2,HILT}. $\Gamma _{i}$ is assumed to depend on the centroid energy through the relation \cite{CARL} $\Gamma _{i}\approx 0.026E_{i}^{1.9}.$ The peak cross section $\sigma _{m}$ is given by \cite{aruprc1}
\begin{equation}
\sigma _{m}=60\frac{2}{\pi }\frac{NZ}{A}\frac{1}{\Gamma }\;(1+\alpha )\;.
\end{equation}
The parameter $\alpha$, which takes care of the sum rule is fixed at 0.3 for all the nuclei. In most of the cases we normalize the peak with the
experimental data and hence the choice of $\alpha $ has a negligible effect on the results. The other parameters $\eta_0$ (or $\eta$) and $\chi_0$ (or
$\chi$) vary with nuclei so that the experimental width of the GDR built
on the ground state is reproduced. The choice for $^{120}$Sn is $\eta_0=2.6$, $\chi_0=3.5$ MeV$^{-5/2}$ and for $^{97}$Tc it is $\eta_0 =2.6$, $\chi_0=1.7$ MeV$^{-5/2}$.
\subsection{Fluctuations}
When the nucleus is observed at finite temperature, the effective GDR
cross-sections carry information on the relative time scales for shape
rearrangements \cite{ALHAT}, which lead to shape fluctuations.  The general expression for the expectation value of an observable $\mathcal{O}$ incorporating  such thermal shape fluctuations is given by \cite{ALHA}\begin{equation}\label{average_1}
\langle \mathcal{O}\rangle _{\beta ,\gamma }=\frac{\int \mathcal{D}[\alpha
]\exp[-F(T;\beta ,\gamma )/T]\mathcal{O}}{\int \mathcal{D}[\alpha
]\exp[-F(T;\beta ,\gamma )/T]}\;
\end{equation}%
with $\mathcal{D}[\alpha ]=\beta ^{4}|\sin 3\gamma |\,d\beta \,d\gamma $. With the inclusion of pairing fluctuations, we have
\begin{equation}\label{average_2}
\langle \mathcal{O}\rangle _{\beta ,\gamma ,\Delta _{P},\Delta _{N}}=\frac{%
\int \mathcal{D}[\alpha ]\exp[-F(T;\beta ,\gamma ,\Delta _{P},\Delta _{N})/T]%
\mathcal{O}}{\int \mathcal{D}[\alpha ]\exp[-F(T;\beta ,\gamma ,\Delta
_{P},\Delta _{N})/T]}\;
\end{equation}%
with $\mathcal{D}[\alpha ]=\beta ^{4}|\sin 3\gamma |\,d\beta \,d\gamma \
\Delta _{P}\,\Delta _{N}\,d\Delta _{P}\,d\Delta _{N}$.
We perform the  TSF calculations exactly by numerically computing the 
integrations in Eq.~(\ref{average_2}) with the free energy and the observables calculated at every mesh point (deformations and pairing gaps), utilizing
the microscopic-macroscopic approach outlined in Sec.~\ref{sec_defene}. 

\section{Results and Discussion}

\begin{figure}
\center\includegraphics[width=0.99\columnwidth]{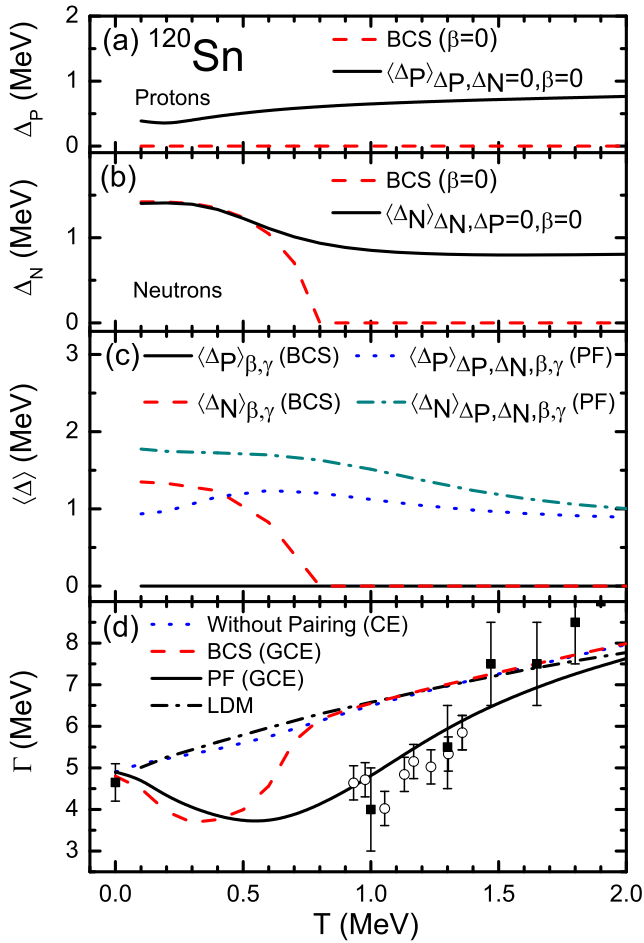}
\caption{(Color online) Plots (a) and (b) represent the pairing gaps as a function of temperature for  $^{120}$Sn with $\beta=0$ calculated from the simple BCS approach and with the pairing fluctuations (without shape fluctuations). Full results for both protons and neutrons are also shown: (c) average pairing gap, (d) GDR width in the case of $^{120}$Sn, as a function of temperature. The calculations  done without pairing utilize  a CE approach (CE), whereas the calculations with  simple BCS pairing (BCS) and with pairing fluctuations (PF) utilize a GCE approach (GCE). The results obtained using the liquid drop model (LDM) are also presented. Experimental values for  $^{120}$Sn are taken from Refs.~\cite{Baum98,Heck43} are shown by solid squares. For comparison, data for $^{119}$Sb, taken from Ref.~\cite{mukhu9}, are also shown with open circles.}
\label{fig1}
\end{figure}

\begin{figure}
\center\includegraphics[width=0.99\columnwidth]{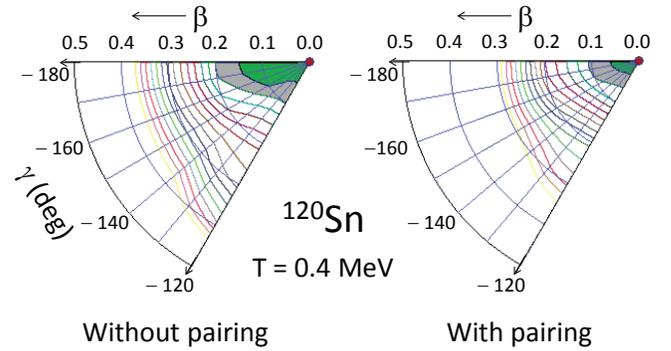}
\caption{(Color online) The free energy surfaces for $^{120}$Sn at $T=0.4$
MeV plotted without pairing and with pairing where the calculations
are with CE and GCE, respectively. The contour line spacing is 0.5 MeV, the most probable shape is marked by a solid (red) circle and the first two minima are shaded.} \label{fig2}
\end{figure}

\begin{figure}
\center\includegraphics[width=0.99\columnwidth]{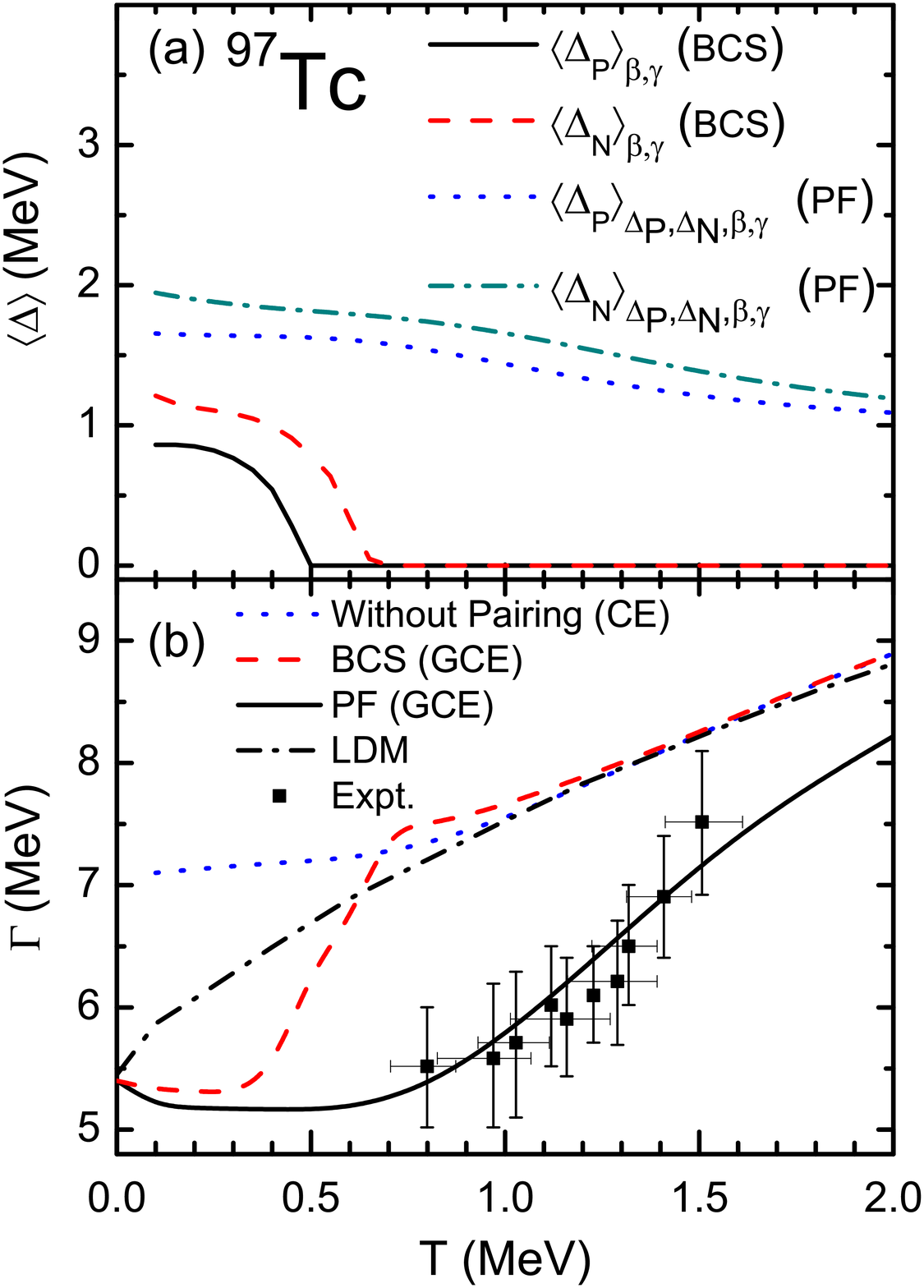}
\caption{(Color online) Similar to   Figs.~\ref{fig1}(c)
and~\ref{fig1}(d) but for the nucleus $^{97}$Tc. Experimental values  are taken from Ref.~\cite{Balaram}   are shown with solid squares.} \label{fig3}
\end{figure}

\begin{figure}
\center\includegraphics[width=0.99\columnwidth]{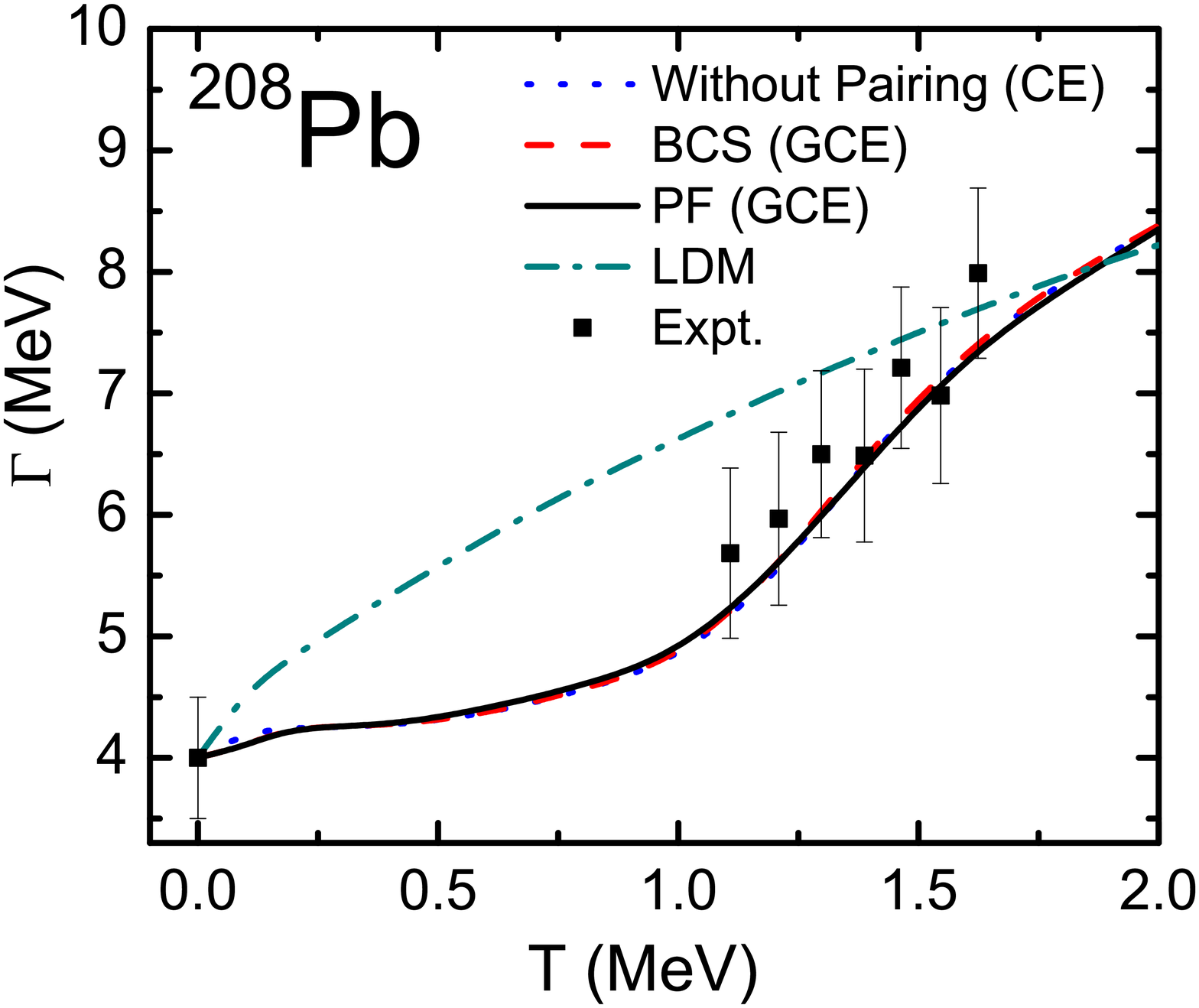}
\caption{(Color online) Similar to Fig.~\ref{fig1}(d) but for the nucleus  $^{208}$Pb.  Experimental values   are taken from Ref.~\cite{Kusn98} are shown with  solid squares. } \label{fig4}
\end{figure}

We categorize our different approaches into the following cases:
\begin{enumerate}
\item \textbf{LDM} - The free energies correspond to that of the liquid drop
model and hence no pairing is included.
\item \textbf{Without pairing (CE)} - The free energies are calculated assuming
a CE and the pairing correlations are neglected.
\item \textbf{BCS (GCE)} - The free energies are calculated assuming
a GCE and the pairing has been taken care of within
the simple BCS approach.
\item \textbf{PF (GCE)} - The free energies are calculated assuming
a GCE and along with the pairing and its fluctuations are
also has been taken care of (\ref{average_2}). The first three cases are with TSF only (\ref{average_1}).
\end{enumerate}

It is convenient to consider the following important factors which can affect
the GDR width ($\Gamma$) in our calculations:
\begin{enumerate}
\item[A.] The shell effects 
through the modification of free energy surfaces.
\item[B.] The pairing effects:
\begin{enumerate}
\item[i.] modification of free energy surfaces.
\item[ii.] damping of the GDR through the term $\chi P^\dagger P~(\ref{GDRhamiltonian}).$
\end{enumerate}
\end{enumerate} 

We start with the study of temperature dependence of GDR width in the case of $^{120}$Sn where the measured values~\cite{Baum98,Heck43} have been used as a benchmark for several theories.  The results from our calculations are presented in Fig.~\ref{fig1}. In Fig.~\ref{fig1}(a) we see that with the inclusion of fluctuations in the pairing field, even for a closed shell, the average pairing gap is finite and also sustained at high
$T$. The trend for $\Delta_N$ (Fig.~\ref{fig1}(b)) is in accordance with earlier works \cite{MORET40B} with the average value converging to the BCS value as $T\rightarrow 0$. Once we include the TSF along with the PF, the average pairing gaps acquire a larger value as seen in Fig.~\ref{fig1}(c). 

The GDR widths obtained within various approaches are presented in Fig.~\ref{fig1}(d).
We observe that the shell effects have a very little role on $\Gamma$ because the results with LDM and those without pairing are almost the same. Shell effects can contribute to change in $\Gamma$ only by modifying the free energy surfaces.  The modification in free energy surface can be of two types viz., 
\begin{enumerate}
\item[a.] Change in the minimum with $\Gamma$ proportional to the quadrupole
deformation ($\beta$) corresponding to the minimum. 

\item[b.] Change in surface around
the minimum: a deeper or crisper minimum will lead to sampling over a smaller
region of deformations (lesser TSF) and a shallow or well-spread minimum will lead to contributions from larger deformations (stronger TSF). With $\Gamma \propto \beta$, lesser TSF leads to a quenched  $\Gamma$.  
\end{enumerate}

The free energy surfaces for $^{120}$Sn at $T=0.4$ MeV are presented in Fig.~\ref{fig2}
where we can see that the one without pairing resembles that
of a LDM and hence the shell effects are subtle.  With pairing, the surface around minimum energy becomes much crisper and leads to a quenched  $\Gamma$. We can notice from Fig.~\ref{fig1}(c) that, while treating pairing within the BCS approach there is no proton pairing and the averaged neutron pairing gap vanishes at $T=0.8$ MeV. With the inclusion of PF, the proton pairing develops, the neutron pairing strengthens and it is sustained even at $T=2$ MeV. Subsequently the quenching of $\Gamma$ is also sustained at higher $T$ and the results agree with the experimental observations.

 Very recently, experimental data at low $T$ for the nuclei $^{201}$Tl~\cite{Pandit2012434} and $^{97}$Tc~\cite{Balaram} are reported and our results for $^{97}$Tc  nucleus is presented here. In $^{97}$Tc, with BCS calculations, the proton pairing gap and neutron pairing gap vanish at $T=0.5$ MeV and $T\sim0.65$ MeV respectively as shown in Fig.~\ref{fig3}(a). While considering the  PF along with TSF, the average pairing gaps continue to be strong even $T=2$ MeV. At low $T$, the  discrepancy between ``LDM" and ``Without pairing" cases are due to the fact that the  shell effects drive the shape of this nucleus
to a  deformed one (and hence a larger $\Gamma$). The inclusion of pairing leads to a deeper minimum and hence  quenched back $\Gamma$ as shown  by
the  dashed line in Fig.~\ref{fig3}(b). 

 However, this quenching in $\Gamma$ owing to BCS pairing is not sufficient to explain the experimental data. The latter can be explained only after
the PF along with TSF are included as shown by  the  solid line in Fig.~\ref{fig3}(b).
Here we have demonstrated that with the inclusion of not just pairing but its fluctuations also is crucial to explain the observed quenching.  

At low $T$, in $^{120}$Sn we have seen that the shell effects have no role
on $\Gamma$ and they increase $\Gamma$ in the case of $^{97}$Tc at very low $T$. In $^{208}$Pb, it is known \cite{aruepl} that the shell effects quench the GDR width as a consequence of a deeper minimum in the free energy surface.
Our results in this case are shown in Fig.~\ref{fig4}, where we see that
the quenching in $\Gamma$ occurs once we include the shell effects. Even for such a doubly magic nucleus, we can see a significant pairing gap with the inclusion of PF. However, the PF has no contribution to $\Gamma$, which is dominated by the shell effects only.  More experimental data at lower $T$ could be useful to validate this argument.

 However, our results for $^{97}$Tc, $^{120}$Sn and $^{208}$Pb clearly  demonstrate that the TSFM can be quite successful, if the shell effects (with explicit temperature dependence) and the pairing effects are properly incorporated in the free energy. To strengthen our arguments, it would be nice to have more experimental data for a single nucleus but at several temperatures along with a precise data at $T=0$.

\section{Conclusions}We have constructed  a theoretical framework to study the GDR with proper treatment of  pairing and its fluctuations along with the thermal shape fluctuations. Our study reveals
that the observed quenching of GDR width at low temperature in the open-shell
nuclei $^{97}$Tc and $^{120}$Sn can be understood in terms of simple shape effects caused by the pairing correlations. For a precise match with the experimental data, the consideration of pairing fluctuations is crucial. 

 Our results for $^{97}$Tc, $^{120}$Sn and $^{208}$Pb clearly  demonstrate that the TSFM can be quite successful if the shell effects (with explicit temperature dependence) and the pairing ones are properly incorporated in the free energy. It has to be noted that often the empirical formula in
Ref.~\cite{Kusn98} (constructed to mimic the results of TSFM at higher temperatures) is quoted as the phenomenological TSFM. The subsequent failure of this empirical
formula as well as its modifications in describing GDR width at low temperature
should not be confused with that of the TSFM (also called the adiabatic TSFM) because of the absence of thermal pairing in the latter.


This  work is supported by the Department of Science
and Technology, Government of India. Project No. SR/FTP/PS-086/2011.
A part of this work was completed at RIKEN and the numerical
calculations were carried out using RIKEN Integrated Cluster of Clusters (RICC).

\end{document}